**Prediction of the 2023 Turkish Presidential Election Results Using Social Media Data**

**Aysun Bozanta*^, Fuad Bayrak*, Ayse Basar***

*Management Information Systems Department, Bogazici University, Istanbul, Turkey
**Data Science Lab, Toronto Metropolitan University, Toronto, Canada
^corresponding author: aysun.bozanta@boun.edu.tr

**Abstract**

Social media platforms influence the way political campaigns are run and therefore they have become an increasingly important tool for politicians to directly interact with citizens. Previous elections in various countries have shown that social media data may significantly impact election results. In this study, we aim to predict the vote shares of parties participating in the 2023 elections in Turkey by combining social media data from various platforms together with traditional polling data. Our approach is a volume-based approach that considers the number of social media interactions rather than content. We compare several prediction models across varying time windows. Our results show that for all time windows, the ARIMAX model outperforms the other algorithms.

**Introduction**

Elections are a cornerstone of democratic societies, providing citizens with the opportunity to have their voices heard and to shape the future of their countries. Predicting the election results is important for both citizens and political parties. Political parties have loyal supporters who do not change their minds, and they do not swing from one party to the other. However, many citizens may be indecisive and/ or vote strategically to influence a particular outcome.  Likewise, predicting the election results have important implications for policy and governance. If the outcome of an election is known in advance, political parties and candidates can adjust their strategies and policies accordingly. This can lead to more effective and responsive governance, as elected officials are better able to anticipate and address the needs and concerns of their constituents.

Polls and surveys are traditional methods used for predicting election results. However, they have various limitations. The first limitations are sample related disadvantages. Traditional polling methods typically rely on relatively small samples of the population, which may not represent the broader electorate. Traditional polling methods rely on voluntary participation, which can lead to non-response bias if certain groups are more likely to participate than others. Traditional polling methods are often conducted through interviews or surveys, which can be subject to social desirability bias which means that people state the socially desirable or acceptable choice, rather than their true opinions. Other limitations can be as follows: they are typically conducted in a relatively short time frame leading up to the election, which may not capture changes in public opinion. They can be expensive and resource-intensive, which may limit their feasibility for smaller campaigns or organizations with limited resources.

The rise of social media has transformed the way people communicate and access news. Social media platforms provide real-time news from a variety of sources, making them a preferred medium for news consumption. In addition, social media allows for direct communication between individuals without the need for intermediaries, making it a more sincere and authentic form of interaction. Politicians have also recognized the potential of social media as a means of communication and interaction with the public. Social media platforms enable politicians to connect with their constituents in real-time and on a more personal level, allowing them to engage with voters and build support for their campaigns. This increased use of social media in political campaigns has also made it an attractive data source for researchers. Social media platforms provide a wealth of data on user behavior, allowing researchers to directly observe and analyze interactions between politicians and voters. As a result, a growing body of literature investigates the relationship between social media data and voter behavior (Tsakalidis, 20155; Brito, 2020; Heredia, 2018; Bovet, 2018). These studies have shown that social media can be a valuable tool for predicting election outcomes and understanding the factors that influence voter decision-making.

In this study, we follow a volume-based approach focusing on the number of posts of political leaders and the number of interactions between the political leaders and citizens. We extend the study of Brito (2020) by adding new algorithms to predict election results in Turkey. The approach is based on modeling and using SM data in a new form focused on the interactions of the posts of official candidates' profiles in all three major SNs (Facebook, Twitter, and Instagram). Then, interaction data were combined with traditional polls and used to train ML models individually for each candidate to predict their vote share. For modeling, we use two different approaches as machine learning algorithms and a statistical model. The contribution of our study can be summarized as follows:

1. To the best of our knowledge, this is the first study that predicts Turkey's election results with a volume-based approach combining two different data types such as social media data from Twitter, Facebook, and Instagram, and the survey data.
2. We used four years' worth of data that includes the daily shares of votes of each candidate. We conducted experiments with various algorithms to predict the daily vote shares of the candidates.
3. In this study, we applied a volume-based approach that focuses on the prediction of the vote shares of political leaders using the number of posts of the political leaders and the interaction count between the political leaders and citizens.

**Literature Review**

The prediction of elections and opinions on political events has garnered significant attention in research (Lewis-Beck, Michael S., 2005). Over the past two decades, political scientists have developed statistical models aimed at predicting future presidential elections (Montgomery, J., Hollenbach, F., & Ward, M.,2012).

With the advent of technological advancements, the utilization of social media data for predicting election results has become increasingly widespread. Two main approaches have emerged: volume-based approaches and content-based approaches. Volume-based approaches involve quantifying tweets, users, or mentions related to a candidate or political party, while content-based approaches focus on sentiment analysis to classify tweets based on expressed agreement or polarity towards a party or candidate (Coletto, 2015; Tung, KC., Wang, E.T., Chen, A.L.P., 2016).

Nonetheless, both approaches possess inherent limitations. Volume-based methods may underperform due to arbitrary choices and inconsistencies across datasets (Allcott, Hunt, and

Matthew Gentzkow, 2017). On the other hand, content-based methods encounter challenges in identifying sarcasm and irony, resulting in inconsistent results across different datasets (Dmitry Davidov, Oren Tsur, and Ari Rappoport, 2010). Additionally, it is vital to consider the bias of Twitter users, as they may not represent the general population or all demographic groups, thereby influencing the predictive power of Twitter (Coletto, 2015).

In addressing the prediction of election outcomes, a novel approach utilizing machine learning models have been proposed (Brito, 2020). This approach focuses on modeling and analyzing the impact of posts from official candidates' profiles on social media platforms such as Facebook, Twitter, and Instagram. Traditional polls are combined with this data and used to train machine learning models like multilayer perceptron artificial neural networks and linear regression, individually for each candidate.

Although social media data offers the potential for predicting attitudes and behaviors due to its scale and richness, a debate persists regarding its usefulness in understanding public opinion (Jungherr, A., Schoen, H., Posegga, O., & Jürgens, P., 2017). Social media data often encounter issues such as incompleteness, noise, lack of structure, unrepresentativeness, and confounding by algorithms. These factors make it less reliable compared to carefully designed and representative surveys (Skoric, 2020).

Moreover, social media-based predictions challenge traditional survey-based research methods. Probability-based sampling, which assumes all opinions are equally valuable, fails to account for interpersonal influence that plays a crucial role in shaping public opinion (Bail, Christopher, 2014). Social media analysis suggests that politically active internet users can act as opinion-makers, influencing the preferences of a wider audience (Skoric, 2020).

Despite the cost-effectiveness and time efficiency of collecting and analyzing public opinion data through social media platforms, particularly Twitter, concerns have been raised about the rigor and reproducibility of studies associating Twitter activity with civic or electoral outcomes (Hilbert, 2019).

Early studies by O'Connor et al. (2010), Mejova et al. (2013), Mitchell and Hitlin (2013), and Ceron et al. (2013) explored the feasibility of using tweets to understand public opinion. These studies employed sentiment analysis methods and yielded mixed results, highlighting issues such as the prevalence of negative sentiment, demographic and self-selection biases, and limited correlation with traditional polls (O'Connor et al., 2010; Mejova et al., 2013; Mitchell and Hitlin, 2013; Ceron et al., 2013).

In conclusion, when it comes to predicting elections and gauging opinions on political events using social media data, particularly on platforms like Twitter, researchers utilize both volume-based and content-based approaches. However, it is crucial to acknowledge the limitations associated with these approaches.

Volume-based methods rely on quantifying metrics such as tweets, users, or mentions related to candidates or political parties. Content-based methods, on the other hand, focus on sentiment analysis to classify tweets based on expressed agreement or polarity. While these methods provide valuable insights, challenges arise from arbitrary choices and inconsistencies in the metric selection, as well as difficulties in accurately identifying sarcasm, irony, and nuanced expressions. Sentiment analysis itself is complex and susceptible to biases.

To enhance prediction accuracy, researchers have proposed machine learning models that combine social media data with traditional polls. However, the usefulness of social media data in understanding public opinion remains a subject of debate. Critics emphasize concerns regarding the incompleteness of social media data, the presence of noise and misinformation, its unrepresentativeness of the general population, and the potential for algorithmic biases.

Moreover, there are apprehensions about the influence of politically active internet users who may not represent the broader population, as well as the reliability of studies linking social media activity to civic or electoral outcomes. These factors underscore the need for caution and critical analysis when utilizing social media data to understand public opinion and make predictions in the political realm.

**Methodology**

In this section, we discuss the datasets, the algorithms that we used, and the evaluation metrics for measuring the performance of the algorithms.

*Datasets*

We used two types of datasets for this study. The first type is traditional survey data collected monthly between November 2019 and May 2023. Türkiye Report surveys are conducted with Computer Assisted Telephone Interviews (CATI). The survey is conducted by Istanbul Ekonomi Research with approximately 2000 people in 26 provinces across Turkey every month[1]. The 26 provinces we included in our research represent 12 NUTS regions of Turkey. Turkey NUTS (Statistical Regional Units Classification) Turkey within the Statistical Regional Units Classification used by European Union countries used for the classification. This standard was developed in 2003 by the European Union. The raw data we obtained as a result of the survey to represent the general population of Turkey are weighted using education, age, and gender information in accordance with the Turkish NUTS standard. The survey results provide statistically significant results with a 95% confidence interval and +/- 2.5 margin of error. The survey data includes the voting percentages of political parties over the years.

The second type is social media data collected daily from Twitter, Facebook, and Instagram. This data shows the number of posts that political party leaders shared on their social media accounts and the amount of interaction the public has with these posts. Table 1 presents the social media platforms, the features from these platforms that we used for the prediction, and their descriptions. Since we collected daily social media data, we interpolated the survey data to make it daily as well as to run daily predictions.

Table 1. Social Media Data Description

| Social Media | Feature | Description |
| --- | --- | --- |
| Twitter | Post | The number of posts that leaders share on Twitter in a given time window |
| | Like | The sum of likes that posts have on Twitter in a given time window |
| | Retweet | The sum of retweets that posts have on Twitter in a given time window |

---

[1] https://turkiyeraporu.com/wp-content/uploads/2023/05/Rapor-85-TR-Mayis-2023-Vol.-I.pdf

|  | Reply | The sum of replies that posts have on Twitter in a given time window |
|---|---|---|
| Facebook | Post | The number of posts that leaders share on Facebook in a given time window |
|  | Like | The sum of likes that posts have on Facebook in a given time window |
|  | Comment | The sum of comments that posts have on Facebook in a given time window |
|  | Shares | The sum of shares that posts have on Facebook in a given time window |
| Instagram | Post | The number of posts that leaders share on Instagram in a given time window |
|  | Like | The sum of likes that posts have on Instagram in a given time window |
|  | Share | The sum of shares that posts have on Instagram in a given time window |

*Algorithms*

We used four different prediction models that are explained below. For all the algorithms that we used, instead of random allocation, we employed a time-series approach to divide our training and test sets, effectively incorporating the temporal dimension. As predictors of vote shares, we leveraged social media interactions.

Linear regression is a supervised learning algorithm that aims to model the relationship between a dependent variable (target variable) and one or more independent variables (predictor variables). It assumes a linear relationship between the variables, attempting to find the best-fitting line that minimizes the difference between the predicted and actual values (James et al., 2017).

Random Forest Regressor is a powerful ensemble learning algorithm used for regression tasks in machine learning. It combines the strength of multiple decision trees to make accurate predictions. The algorithm constructs a multitude of decision trees using random subsets of the training data and features and then aggregates the predictions of these trees to produce the final output. Each tree independently learns a portion of the input space, reducing overfitting and improving generalization. By considering the average or majority vote of the individual tree predictions, Random Forest Regressor can handle nonlinear relationships, handle missing values, and provide robustness against outliers (Breiman, 2001). Gradient Boosting Regressor is a powerful machine learning algorithm used for regression tasks that iteratively builds an ensemble of weak prediction models to create a strong predictive model. It works by initially fitting a simple model to the data and then sequentially adding new models, each one attempting to correct the mistakes made by the previous models. The algorithm assigns weights to the data points based on their error residuals, with greater emphasis on the points that were poorly predicted. The new models are trained to minimize the residual errors of the previous models. By combining the predictions from multiple weak models, Gradient Boosting Regressor creates a highly accurate and robust predictive model capable of capturing complex relationships in the data (Freidman, 2011).

ARIMAX (AutoRegressive Integrated Moving Average with Exogenous Variables) is a time series forecasting model that extends the traditional ARIMA (AutoRegressive Integrated Moving Average) model by incorporating exogenous variables. It is used to analyze and predict time-dependent data that may be influenced by external factors. ARIMAX takes into account both the autoregressive (AR) and moving average (MA) components to capture the dependencies within the time series, while also considering the impact of exogenous variables on the series. By incorporating these additional variables, ARIMAX can account for external factors that may affect the behavior of the time series, leading to more accurate and robust predictions. This makes ARIMAX a valuable tool in various applications such as economic forecasting, demand prediction, and financial analysis, where external factors can significantly influence the time series being analyzed (Box et al., 2015).

*Evaluation Metrics*

We used two evaluation metrics namely Root Mean Square Error (RMSE) and Mean Absolute Error (MAE).

RMSE is a widely used metric for evaluating the accuracy of a predictive model, particularly in regression tasks. It measures the average magnitude of the differences between the predicted values and the actual values of a dataset. RMSE is calculated by taking the square root of the mean of the squared differences between the predicted and actual values. It provides a single value that represents the typical deviation or error of the model's predictions from the true values. A lower RMSE indicates a better fit between the predicted and actual values, with zero, indicating a perfect match (Willmott & Matsuura, 2005).

MAE is a commonly used metric for evaluating the accuracy of a predictive model, particularly in regression tasks. It measures the average magnitude of the differences between the predicted values and the actual values of a dataset. MAE is calculated by taking the mean of the absolute differences between the predicted and actual values. Unlike RMSE, MAE does not square the errors, which makes it more robust to outliers (Witkowski & Rutkowski, 2018).

**Results**

In this section, we present the descriptive analysis of data sets and the experimental results comparing the performance of the prediction models for 10 different time windows.

We collected the candidates' posts on Twitter, Facebook, and Instagram between Nov 1, 2019, and Jan 1, 2023. Tables 2 and 3 present the number of posts that candidates shared during this time and their interactions on social media platforms. We consider only Kemal Kılıçdaroğlu and Recep Tayyip Erdoğan in this study since they are the major candidates by taking more than 90% of the votes.

Table 2. The number of posts of Kemal Kılıcdaroglu and their interactions

| Social Media | Feature | Min | Max | Mean | Std. Dev. |
|---|---|---|---|---|---|
| Twitter | Post | 1 | 22 | 2.18 | 1.62 |
| | Like | 1718 | 450,026 | 45,405.50 | 53,972.29 |
| | Retweet | 183 | 65,902 | 5,160.18 | 6,336.01 |
| | Reply | 114 | 29,884 | 3,377.89 | 3,927.55 |
| | Post | 1 | 7 | 1.86 | 1.06 |
| Facebook | Like | 1200 | 167,000 | 16,372.40 | 15,350.86 |
| | Comment | 46 | 27,500 | 2,606.20 | 2,880.76 |
| | Shares | 41 | 12,100 | 1,876.02 | 1,657.93 |
| Instagram | Post | 0 | 5 | 1.29 | 0.59 |
| | Like | 0 | 4,724,120 | 485,787.72 | 426,963.44 |
| | Share(comment) | 2,310 | 179,820 | 19,601.68 | 20,457.15 |

Table 3. The number of posts of Recep Tayyip Erdoğan and their interactions

| Social Media | Feature | Min | Max | Mean | Std. Dev. |
|---|---|---|---|---|---|
| Twitter | Post | 1 | 257 | 4.3 | 7.23 |
| | Like | 9 | 3,241,605 | 61,702.06 | 116,693 |
| | Retweet | 16 | 1,317,198 | 17,086.71 | 40,882.07 |
| | Reply | 1 | 110,776 | 4,248.32 | 7,548.85 |
| Facebook | Post | 1 | 18 | 3.18 | 2.47 |
| | Like | 3500 | 689,000 | 72,004.91 | 74,402.03 |
| | Comment | 46 | 125,000 | 6,895.86 | 8,897.32 |
| | Shares | 59 | 89,900 | 4,633.98 | 6,316.15 |
| Instagram | Post | 1 | 10 | 1.73 | 1.44 |
| | Like | 29,169 | 1,969,447 | 333,650.42 | 292,143.55 |
| | Share (comment) | 0 | 191,601 | 6,883.42 | 11,296.15 |

The poll data is collected between Nov 2019 and May 2023. Until April 2023, the survey data were collected monthly. As the election time gets closer the survey data is collected in 2 or 3-week periods starting in April 2023. Figures 1 and 2 present the seasonal decomposition of poll data for vote shares of candidates; Kemal Kılıcdaroglu and Recep Tayyip Erdoğan respectively.

Figure 1. Seasonal Decomposition of the vote share of Kemal Kılıcdaroglu

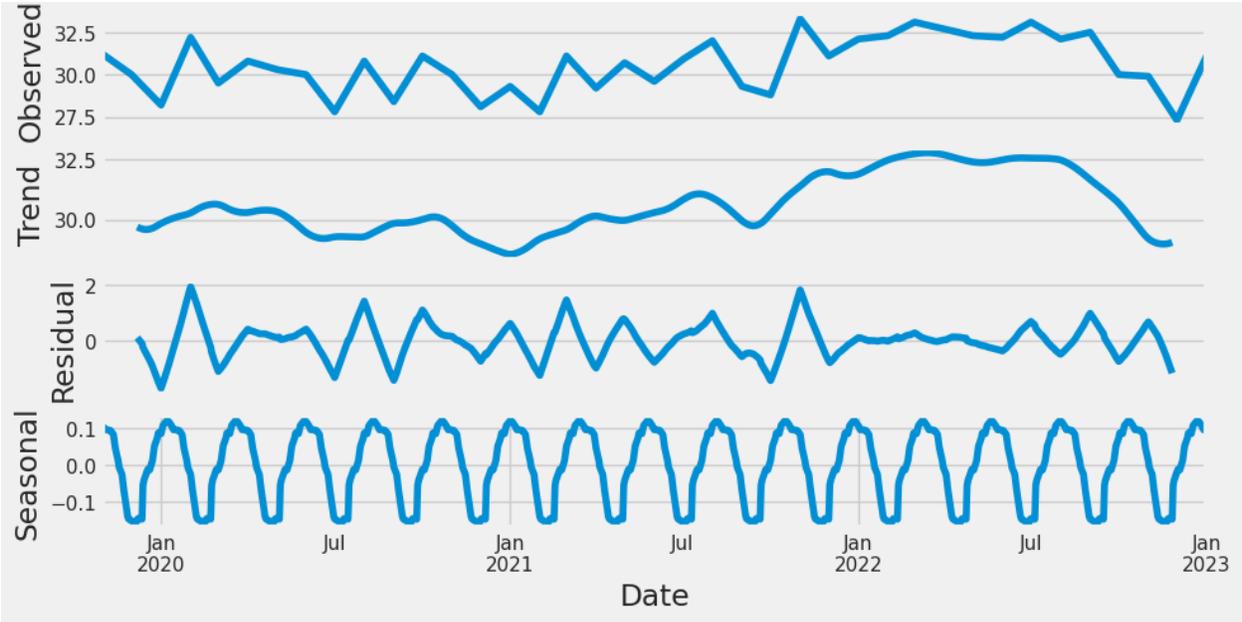

Figure 2. Seasonal Decomposition of the vote share of Recep Tayyip Erdoğan

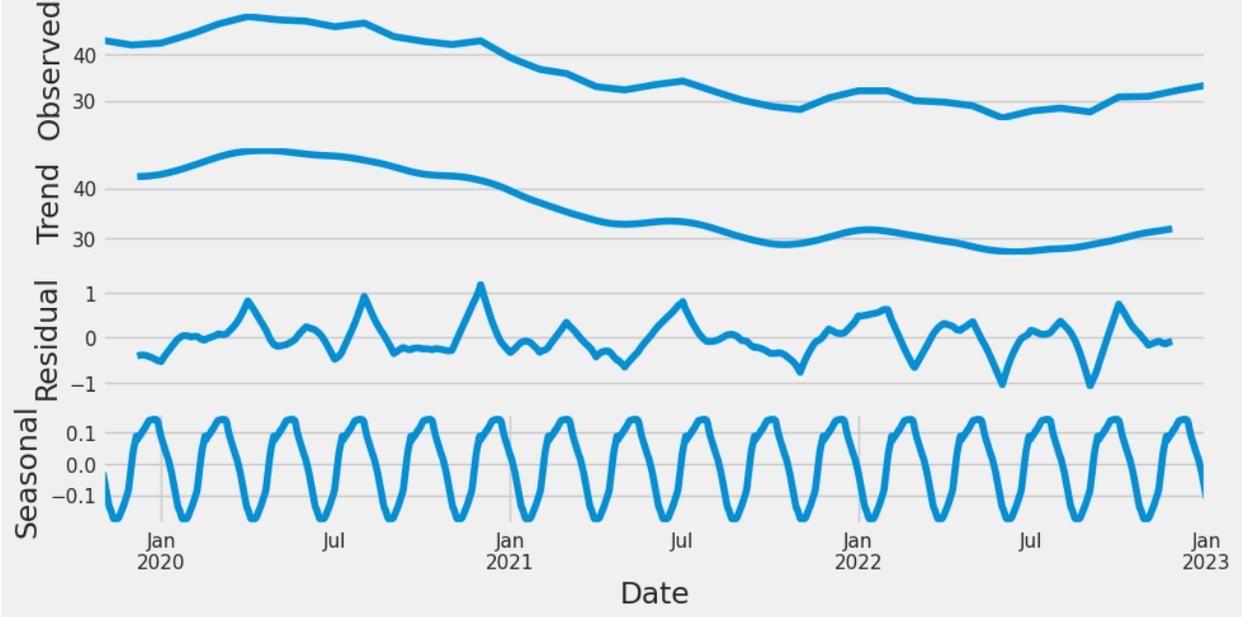

Figure 3 presents the vote shares of the political parties through the years. We interpolated monthly poll data and made it daily as well since we collected the social media data daily.

Figure 3. The vote shares of the political parties based on the poll data

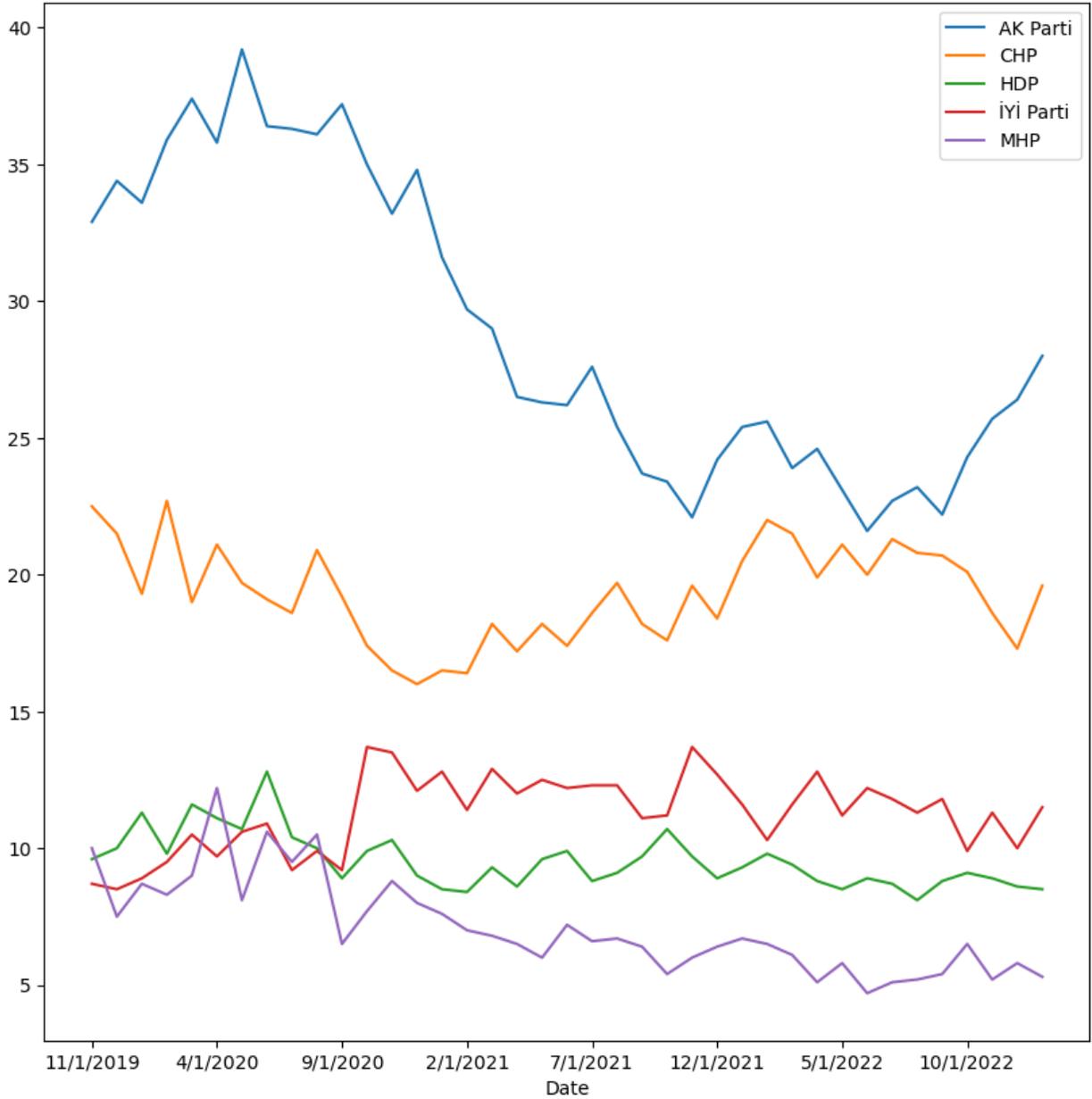

A set of 10 independent datasets is generated for each of the five candidates considered, with the features explained in Table 1. Each dataset is generated with a different aggregated arbitrary window, w = [1, 2, 3, 4, 5, 6, 7, 14, 21, 28], to avoid bias in the experiments. For testing the prediction models over social media posts and vote shares datasets, we apply an augmented out-of-sampling method, which enables updating/retraining the model after each test sample prediction

(Ilic, 2020). In this technique, after each test, the prediction model is updated to include the test data for training purposes. Typically, a sliding window approach should be used when old data is disposed of. In addition, we chose ARIMAX parameters as follows: p is 0, q is 1, and d is 5. The reason for choosing the p values as 0 is to understand the effects of exogenous variables (social media features) on the prediction of vote shares.

We compared the performances of the prediction models for ten prediction windows using different social media data to determine the algorithm, time window, and data set combination based on obtaining the lowest error value.

Table 4. MAE values of the algorithms for changing time windows using different datasets

| Data sets | Time Window | ARIMAX | Linear Regression | Random Forest | Gradient Boosting |
|---|---|---|---|---|---|
| Twitter | 1 | 0,01 | 1,71 | 1,61 | 1,64 |
| | 2 | 0,02 | 1,71 | 1,61 | 1,64 |
| | 3 | 0,02 | 1,71 | 1,61 | 1,65 |
| | 4 | 0,04 | 1,71 | 1,61 | 1,65 |
| | 5 | 0,05 | 1,71 | 1,61 | 1,65 |
| | 6 | 0,05 | 1,71 | 1,62 | 1,66 |
| | 7 | 0,09 | 1,72 | 1,63 | 1,66 |
| | 14 | 0,18 | 1,72 | 1,63 | 1,67 |
| | 21 | 0,28 | 1,73 | 1,64 | 1,67 |
|          | 28 | 0,34 | 1,74 | 1,66 | 1,69 |
| Facebook | 1  | 0,01 | 1,78 | 1,74 | 1,72 |
|          | 2  | 0,02 | 1,78 | 1,75 | 1,73 |
|          | 3  | 0,02 | 1,78 | 1,75 | 1,73 |
|          | 4  | 0,04 | 1,78 | 1,75 | 1,73 |
|          | 5  | 0,05 | 1,78 | 1,75 | 1,73 |
|          | 6  | 0,05 | 1,79 | 1,75 | 1,74 |
|          | 7  | 0,09 | 1,79 | 1,74 | 1,73 |
|          | 14 | 0,18 | 1,80 | 1,76 | 1,75 |
|          | 21 | 0,27 | 1,80 | 1,77 | 1,75 |
|          | 28 | 0,34 | 1,81 | 1,79 | 1,76 |
| Instagram | 1 | 0,01 | 1,79 | 1,76 | 1,76 |
|          | 2  | 0,02 | 1,79 | 1,76 | 1,76 |
|          | 3  | 0,02 | 1,79 | 1,76 | 1,77 |
|          | 4  | 0,04 | 1,79 | 1,76 | 1,77 |

|  | 5 | 0,05 | 1,80 | 1,77 | 1,77 |
|---|---|---|---|---|---|
|  | 6 | 0,05 | 1,80 | 1,76 | 1,77 |
|  | 7 | 0,09 | 1,80 | 1,76 | 1,77 |
|  | 14 | 0,18 | 1,81 | 1,78 | 1,79 |
|  | 21 | 0,27 | 1,81 | 1,79 | 1,80 |
|  | 28 | 0,34 | 1,82 | 1,80 | 1,81 |
| Twitter + Facebook + Instagram | 1 | 0,01 | 1,69 | 1,65 | 1,65 |
|  | 2 | 0,02 | 1,69 | 1,66 | 1,66 |
|  | 3 | 0,02 | 1,70 | 1,65 | 1,66 |
|  | 4 | 0,04 | 1,70 | 1,65 | 1,66 |
|  | 5 | 0,05 | 1,70 | 1,66 | 1,66 |
|  | 6 | 0,04 | 1,70 | 1,67 | 1,67 |
|  | 7 | 0,09 | 1,70 | 1,67 | 1,66 |
|  | 14 | 0,18 | 1,72 | 1,68 | 1,69 |
|  | 21 | 0,27 | 1,73 | 1,69 | 1,69 |

|  | 28 | 0,34 | 1,74 | 1,70 | 1,71 |

Table 4 is prepared to show the performances of the algorithms to predict one of the candidates. Since all algorithms perform very similarly for another candidate, we only present one of the results. In Table 4, we can observe that the ARIMAX model outperforms all other algorithms. The second best-performing algorithms are Random Forest and Gradient Boosting regressors. We obtained the best results when the prediction window is 1 day. There is no significant difference between the performances of the algorithms when we used different datasets and even if we use all the features from three different social media platforms, the results do not get better. After this analysis, we chose our time window as 1 day and continued with Twitter data since there is no difference between using different datasets and it is easy to collect the Twitter data.

Table 5. Actual, Predicted, and Poll Votes Shares of Turkish President Candidates for Round 1

| Candidate | Vote Shares (%) (May 14, 2023) | Vote Shares from Poll (%) | Predicted Vote Shares before adding indecisive voters (%) | Predicted Vote Shares after adding indecisive voters (%) |
|---|---|---|---|---|
| Kemal Kılıcdaroglu | 45.1 | 46.5 | 44.0 | 48.1 |
| Recep Tayyip Erdogan | 49.2 | 44.0 | 42.0 | 45.9 |

We continued to collect Twitter data until May 11, 2023. The last poll data showing the vote shares of candidates and the political parties was collected on May 9, 2023. Table 5 shows the actual vote percentages coming from the election result, the vote percentages coming from the poll data, and the predicted vote percentages by our model. Kemal Kılıçdaroğlu got 45.1% of the votes, while Recep Tayyip Erdoğan got 49.2% of the votes in the election that is took place in May 14, 2023. There are two more candidates namely Sinan Oğan who got 5.2% of the votes and Muharrem İnce who got less than 1% of the votes. We made our prediction before distributing the indecisive voters and then we distributed indecisive voters based on the predicted vote shares. Although our predictions are consistent with the poll data, neither poll data nor our predictions are aligned with the actual election results for the Presidency.

Table 6. Actual, Predicted, and Poll Votes Shares of Turkish Political Parties

| Candidate | Vote Shares (%) (May 14, 2023) | Vote Shares from Poll (%) | Predicted Vote Shares (%) |
|---|---|---|---|
| AKP | 35.6 | 36.7 | 34.9 |
| CHP | 25.4 | 30.6 | 30.0 |
| Iyi Party | 9.7 | 10.1 | 9.6 |
| MHP | 10.1 | 7.4 | 9.0 |
| Yesil Sol Parti | 8.8 | 9.1 | 9.0 |

Table 6 presents the actual vote shares coming from the election result, the vote shares coming from the poll data, and the predicted vote shares by our model for Turkish political parties. For the political parties' vote shares, we made very accurate predictions and they are aligned with the poll data as well. The worst-predicted vote share belongs to the CHP. Although we have made an estimation quite close to the survey data here, a deviation from the actual election result is observed.

Since any of the candidates could get more than 50% of the votes, there will be a second round for the presidential elections on May 28, 2023. As of now, there is no existing poll data for the second round, and the actual results are still undisclosed. Table 7 presents the prediction of the election results of the second round based on different scenarios. To generate these scenarios, we allocated the votes of Sinan Oğan, who received 5.2% of the votes in the initial round, along with votes from undecided voters, in different manners, resulting in a total of 9 scenarios.

Table 7. Predicted Results of Round 2 for Different Scenarios

|  | Kemal Kılıçdaroğlu | Recep Tayyip Erdoğan |
|---|---|---|
| Scenario A | 45.0% | 46.3% |
| Scenario B | 45.0% | 55% |
| Scenario C | 48.5% | 51.5% |
| Scenario D | 50.2% | 49.8% |
| Scenario E | 53.7% | 46.3% |
| Scenario F | 47.6% | 52.4% |
| Scenario G | 51.1% | 48.9% |
| Scenario H | 46.8% | 53.2% |
| Scenario I | 51.9% | 48.1% |
| Scenario J | 49.4% | 50.6% |

The explanations of the scenarios are given below.

Scenario A: Excluding the votes of Sinan Ogan and indecisive voters

Scenario B: Adding the votes of Sinan Ogan and indecisive voters to Recep Tayyip Erdoğan

Scenario C: Adding the votes of indecisive voters to Kemal Kılıçdaroğlu and the votes of Sinan Oğan to Recep Tayyip Erdoğan

Scenario D: Adding the votes of Sinan Oğan to Kemal Kılıçdaroğlu and the votes of indecisive voters to Recep Tayyip Erdoğan

Scenario E: Adding the votes of Sinan Ogan and indecisive voters to Kemal Kılıçdaroğlu

Scenario F: Distributing the votes of Sinan Ogan equally to the candidates and the votes of indecisive voters to Recep Tayyip Erdoğan

Scenario G: Distributing the votes of Sinan Ogan equally to the candidates and the votes of indecisive voters to Kemal Kılıçdaroğlu

Scenario H: Distributing the votes of indecisive voters equal to the candidates and the votes of Sinan Oğan to Recep Tayyip Erdoğan

Scenario I: Distributing the votes of indecisive voters equal to the candidates and the votes of Sinan Oğan to Kemal Kılıçdaroğlu

Scenario J: Distributing the votes of Sinan Oğan and indecisive voters equal to the candidates

The vote percentage range of Kemal Kılıçdaroğlu changes between 45% and 53.7%, while the vote percentage range of Recep Tayyip Erdoğan changes between 46.3% and 55%. The average vote percentages for these scenarios are 49.4% for Kemal Kılıçdaroğlu and 50.6% for Recep Tayyip Erdoğan.

**Discussion**

The results of our study indicate that the ARIMAX model outperforms other algorithms in predicting the vote shares of political parties participating in the 2023 elections in Turkey. Our volume-based approach, which focuses on the number of social media interactions rather than content, proved to be effective in capturing the dynamics of voter behavior.

The use of social media data for predicting election outcomes has gained significant attention in recent years. Previous studies have shown that social media can be a valuable tool for understanding voter behavior and predicting election results (Tsakalidis, 2015; Brito, 2020; Heredia, 2018; Bovet, 2018). Our study contributes to this body of literature by combining social media data from multiple platforms, including Twitter, Facebook, and Instagram, with traditional polling data to predict the vote shares of parties in the Turkish elections.

Traditional polling methods have several limitations, including small sample sizes, non-response bias, and social desirability bias. In contrast, social media platforms provide a wealth of data on user behavior, allowing researchers to directly observe and analyze interactions between politicians and voters. By leveraging this data, we aimed to overcome the limitations of traditional polling methods and improve the accuracy of election predictions.

Our study focused on the volume-based approach, considering the number of posts of political leaders and the interaction count between political leaders and citizens. This approach differs from content-based approaches that rely on sentiment analysis to classify tweets based on expressed agreement or polarity. While sentiment analysis can be informative, it faces challenges in accurately identifying sarcasm, irony, and nuanced expressions, which can lead to inconsistent results across different datasets.

The evaluation of our prediction models using RMSE and MAE metrics demonstrated the superiority of the ARIMAX model. ARIMAX is a time series forecasting model that incorporates exogenous variables, making it suitable for analyzing and predicting time-dependent data influenced by external factors. By considering both autoregressive and moving average components, as well as the impact of exogenous variables, ARIMAX captures the dependencies within the time series and provides more accurate and robust predictions.

It is important to acknowledge the limitations associated with the use of social media data for predicting election outcomes. Social media data may suffer from issues such as incompleteness, noise, lack of structure, unrepresentativeness, and confounding by algorithms (Jungherr et al., 2017; Skoric, 2020). Moreover, social media users may not represent the general population or all demographic groups, introducing potential biases into the predictions. The influence of politically active internet users on shaping public opinion also needs to be considered (Bail, 2014).

Although social media data offers cost-effectiveness and time efficiency compared to traditional surveys, caution is necessary for interpreting and generalizing the findings. The rigorousness and reproducibility of studies linking Twitter activity to civic or electoral outcomes have been questioned (Hilbert, 2019). Therefore, a critical analysis of social media data is essential to ensure reliable and valid predictions.

While our predictions closely aligned with the vote shares of political parties, accurately forecasting the vote shares of presidential candidates proved challenging. Several factors could contribute to this outcome. Firstly, the reliability and validity of social media data, as discussed earlier, may have impacted our accuracy. Secondly, the quality of poll data, even if collected professionally, might not have captured the perspectives of all citizens in a representative manner. Lastly, the announced election results have become a subject of ongoing debate and disagreement. In various cities, the dissident party has voiced objections and raised concerns regarding alleged irregularities and voting discrepancies. However, given the limited timeframe between the first and second rounds, the election results are often accepted as they are.

**Conclusion**

In this study, a novel method was introduced for training machine learning models to predict vote share. The approach involved utilizing social media data obtained from the posts on candidates' official profiles, in combination with traditional polls. The data extracted from candidates' posts were processed and improved by applying aggregation windows and creating new variables, including metrics such as the number of likes, shares, and comments per post. Subsequently, prediction models, namely linear regression, random forest regressor, gradient boosting regressor, and ARIMAX were trained using this enhanced dataset. The ultimate objective was to forecast the outcomes of the upcoming presidential elections in Turkey.

The proposed approach demonstrated highly accurate vote share predictions, as evidenced by the favorable results obtained from the MAE and RMSE error metrics. Remarkably, these promising outcomes were achieved despite the limited availability of polls. Furthermore, these findings align with the conclusions drawn by Bristo (2020) in a similar study. Specifically, the employment of the ARIMAX model with a prediction window of 1 day yielded the most optimal results in our analysis.

The approach introduces innovative elements by individually training each candidate, accommodating diverse online supporter behaviors, and effectively handling the influence of bots. Additionally, it addresses the common issue of bias arising from arbitrary keyword selection and time intervals for search. Notably, this study represents the first known endeavor to validate machine learning models specifically for predicting the 2023 Turkish election.

The study also recognized several challenges that can guide future work. Firstly, there is a need to carefully consider the selection of input polls, as it can potentially impact the results. Secondly, the infrequent availability of poll data and the use of linear interpolation techniques to fill in gaps may introduce certain effects on the outcomes. Moreover, there is room for enhancing ML algorithms by exploring alternative approaches specifically tailored for handling small training sets. Emphasizing these aspects can lead to valuable improvements in the overall accuracy and effectiveness of the predictive models.